%% file: main.tex
\pdfoutput=1

\documentclass[12pt,a4paper]{article}

\usepackage[colorinlistoftodos]{todonotes}

\usepackage{ifthen} 
\newboolean{pdflatex}
\setboolean{pdflatex}{true} 

\newboolean{articletitles}
\setboolean{articletitles}{true} 

\newboolean{uprightparticles}
\setboolean{uprightparticles}{false} 

\newboolean{inbibliography}
\setboolean{inbibliography}{false} 

\input{macros}

\input{preamble}
\usepackage{booktabs}

\begin{document}

\renewcommand{\thefootnote}{\fnsymbol{footnote}}
\setcounter{footnote}{1}

\input{title-LHCb-PAPER}


\renewcommand{\thefootnote}{\arabic{footnote}}
\setcounter{footnote}{0}

\pagestyle{plain} 
\setcounter{page}{1}
\pagenumbering{arabic}


\input{body}
\input{acknowledgements}

\addcontentsline{toc}{section}{References}
\setboolean{inbibliography}{true}
\bibliographystyle{LHCb}
\bibliography{main,standard,LHCb-PAPER,LHCb-CONF,LHCb-DP,LHCb-TDR}


\newpage
\input{LHCb_Authorship_flat_28-Aug-2018/LHCb_Authorship_28-Aug-2018}

\end{document}

%% file: macros.tex
\newcommand{\bdzmux}{\mbox{\ensuremath{\Bb\to\Dz\mu^-\neumb X}}\xspace}
\newcommand{\bdmux}{\mbox{\ensuremath{\Bb\to\Dp\mu^-\neumb X}}\xspace}

\newcommand{\dkkpi}{\mbox{\ensuremath{\Dp\to K^+K^-\pi^+}}\xspace}
\newcommand{\dkpipi}{\mbox{\ensuremath{\Dp\to K^-\pi^+\pi^+}}\xspace}
\newcommand{\dzkpi}{\mbox{\ensuremath{\Dz\to K^-\pi^+}}\xspace}
\newcommand{\dzkk}{\mbox{\ensuremath{\Dz\to K^+K^-}}\xspace}
\newcommand{\dzpipi}{\mbox{\ensuremath{\Dz\to \pi^+\pi^-}}\xspace}
\newcommand{\dzhh}{\mbox{\ensuremath{\Dz\to h^+h^-}}\xspace}

\newcommand{\deltag}{\ensuremath{\Delta_\Gamma}\xspace}

\newcommand{\resultycpkpc}{\ensuremath{0.63}\xspace}
\newcommand{\errstatonlyycpkpc}{\ensuremath{0.15}\xspace}
\newcommand{\errsystycpkpc}{\ensuremath{0.11}\xspace}
\newcommand{\resultdgk}{\ensuremath{0.0153}\xspace}

\newcommand{\errstatonlydgk}{\ensuremath{0.0036}\xspace}
\newcommand{\errsystdgk}{\ensuremath{0.0027}\xspace}

\newcommand{\resultycppipc}{\ensuremath{0.38}\xspace}
\newcommand{\errstatonlyycppipc}{\ensuremath{0.28}\xspace}
\newcommand{\errsystycppipc}{\ensuremath{0.15}\xspace}
\newcommand{\resultdgpi}{\ensuremath{0.0093}\xspace}

\newcommand{\errstatonlydgpi}{\ensuremath{0.0067}\xspace}
\newcommand{\errsystdgpi}{\ensuremath{0.0038}\xspace}

\newcommand{\resultycppc}{\ensuremath{0.57}\xspace}
\newcommand{\errstatycppc}{\ensuremath{0.13}\xspace}
\newcommand{\errsystycppc}{\ensuremath{0.09}\xspace}

\newcommand{\mypaperversion}{}
\newcommand{\mydate}{October 16, 2018}
\newcommand{\mylhcbpapernumber}{LHCb-PAPER-2018-038}
\newcommand{\mycernpapernumber}{CERN-EP-2018-270}

\def\paperauthors{LHCb collaboration} 
\def\paperasciititle{Measurement of the charm-mixing parameter yCP} 
\def\papertitle{Measurement of the charm-mixing parameter \ycp} 
\def\paperkeywords{{High Energy Physics}, {LHCb}} 
\def\papercopyright{\the\year\ CERN for the benefit of the LHCb collaboration} 
\def\paperlicence{CC-BY-4.0 licence}
\def\paperlicenceurl{https://creativecommons.org/licenses/by/4.0/}

%% file: preamble.tex
\usepackage[top=1in, bottom=1.25in, left=1in, right=1in]{geometry}

\usepackage{microtype}
\usepackage{lineno}  
\usepackage{xspace} 
\usepackage{caption} 

\usepackage{graphicx}  
\usepackage{color}
\usepackage{colortbl}
\graphicspath{{./figs/}} 
\DeclareGraphicsExtensions{.pdf,.PDF,png,.PNG}

\usepackage{amsmath} 
\usepackage{amssymb}
\usepackage{amsfonts}
\usepackage{upgreek} 

\newcommand*\patchAmsMathEnvironmentForLineno[1]{%
\expandafter\let\csname old#1\expandafter\endcsname\csname #1\endcsname
\expandafter\let\csname oldend#1\expandafter\endcsname\csname
end#1\endcsname
 \renewenvironment{#1}%
   {\linenomath\csname old#1\endcsname}%
   {\csname oldend#1\endcsname\endlinenomath}%
}
\newcommand*\patchBothAmsMathEnvironmentsForLineno[1]{%
  \patchAmsMathEnvironmentForLineno{#1}%
  \patchAmsMathEnvironmentForLineno{#1*}%
}
\AtBeginDocument{%
\patchBothAmsMathEnvironmentsForLineno{equation}%
\patchBothAmsMathEnvironmentsForLineno{align}%
\patchBothAmsMathEnvironmentsForLineno{flalign}%
\patchBothAmsMathEnvironmentsForLineno{alignat}%
\patchBothAmsMathEnvironmentsForLineno{gather}%
\patchBothAmsMathEnvironmentsForLineno{multline}%
\patchBothAmsMathEnvironmentsForLineno{eqnarray}%
}

\usepackage{hyperxmp}

\usepackage[pdftex,
            pdfauthor={\paperauthors},
            pdftitle={\paperasciititle},
            pdfkeywords={\paperkeywords},
            pdfcopyright={Copyright (C) \papercopyright},
            pdflicenseurl={\paperlicenceurl}]{hyperref}

\usepackage[all]{hypcap} 

\input{lhcb-symbols-def} 

\usepackage{cite} 
\usepackage{mciteplus}

%% file: lhcb-symbols-def.tex
\usepackage{xspace} 
\usepackage{upgreek}

\def\lhcb   {\mbox{LHCb}\xspace}

\def\MagUp {\mbox{\em Mag\kern -0.05em Up}\xspace}

\ifthenelse{\boolean{uprightparticles}}%
{

 \def\Pnu         {\ensuremath{\upnu}\xspace}

 \def\PDelta      {\ensuremath{\Delta}\xspace}                 
 \def\PXi      {\ensuremath{\Xi}\xspace}                 
 \def\PLambda      {\ensuremath{\Lambda}\xspace}                 
 \def\PSigma      {\ensuremath{\Sigma}\xspace}                 
 \def\POmega      {\ensuremath{\Omega}\xspace}                 
 \def\PUpsilon      {\ensuremath{\Upsilon}\xspace}                 
 
 \def\PB      {\ensuremath{\mathrm{B}}\xspace}                 
                  
 \def\PD      {\ensuremath{\mathrm{D}}\xspace}

 \def\PK      {\ensuremath{\mathrm{K}}\xspace}

 \def\Pb      {\ensuremath{\mathrm{b}}\xspace}

 \def\Pi      {\ensuremath{\mathrm{i}}\xspace}

 \def\Pp      {\ensuremath{\mathrm{p}}\xspace}

}
{

 \def\Pnu         {\ensuremath{\nu}\xspace}

 \mathchardef\PDelta="7101
 \mathchardef\PXi="7104
 \mathchardef\PLambda="7103
 \mathchardef\PSigma="7106
 \mathchardef\POmega="710A
 \mathchardef\PUpsilon="7107
                  
 \def\PB      {\ensuremath{B}\xspace}                 
                  
 \def\PD      {\ensuremath{D}\xspace}

 \def\PK      {\ensuremath{K}\xspace}

 \def\Pb      {\ensuremath{b}\xspace}

 \def\Pi      {\ensuremath{i}\xspace}

 \def\Pp      {\ensuremath{p}\xspace}

}

\makeatletter
\ifcase \@ptsize \relax
  \newcommand{\miniscule}{\@setfontsize\miniscule{4}{5}}
\or
  \newcommand{\miniscule}{\@setfontsize\miniscule{5}{6}}
\or
  \newcommand{\miniscule}{\@setfontsize\miniscule{5}{6}}
\fi
\makeatother

\DeclareRobustCommand{\optbar}[1]{\shortstack{{\miniscule (\rule[.5ex]{1.25em}{.18mm})}
  \\ [-.7ex] $#1$}}

\def\neu        {{\ensuremath{\Pnu}}\xspace}
\def\neub       {{\ensuremath{\overline{\Pnu}}}\xspace}

\def\neumb      {{\ensuremath{\neub_\mu}}\xspace}
\def\neut       {{\ensuremath{\neu_\tau}}\xspace}
\def\neutb      {{\ensuremath{\neub_\tau}}\xspace}

\def\bquark    {{\ensuremath{\Pb}}\xspace}

  \def\Kbar    {{\kern 0.2em\overline{\kern -0.2em \PK}{}}\xspace}

\def\KorKbar    {\kern 0.18em\optbar{\kern -0.18em K}{}\xspace}

  \def\Dbar    {{\kern 0.2em\overline{\kern -0.2em \PD}{}}\xspace}
\def\D       {{\ensuremath{\PD}}\xspace}

\def\DorDbar    {\kern 0.18em\optbar{\kern -0.18em D}{}\xspace}
\def\Dz      {{\ensuremath{\D^0}}\xspace}
\def\Dzb     {{\ensuremath{\Dbar{}^0}}\xspace}
\def\Dp      {{\ensuremath{\D^+}}\xspace}

\def\B       {{\ensuremath{\PB}}\xspace}
\def\Bbar    {{\ensuremath{\kern 0.18em\overline{\kern -0.18em \PB}{}}}\xspace}
\def\Bb      {{\ensuremath{\Bbar}}\xspace}
\def\BorBbar    {\kern 0.18em\optbar{\kern -0.18em B}{}\xspace}

\def\Bzb     {{\ensuremath{\Bbar{}^0}}\xspace}

\def\Bub     {{\ensuremath{\B^-}}\xspace}

  \def\Y#1S{\ensuremath{\PUpsilon{(#1S)}}\xspace}

\def\proton      {{\ensuremath{\Pp}}\xspace}

\def\Lbar        {{\ensuremath{\kern 0.1em\overline{\kern -0.1em\PLambda}}}\xspace}
\def\LorLbar    {\kern 0.18em\optbar{\kern -0.18em \PLambda}{}\xspace}

\def\to                 {\ensuremath{\rightarrow}\xspace}

\def\CP                {{\ensuremath{C\!P}}\xspace}

\def\AT#1     {\ensuremath{A_{\mathrm{T}}^{#1}}\xspace}           

\def\C#1      {\ensuremath{\mathcal{C}_{#1}}\xspace}                       
\def\Cp#1     {\ensuremath{\mathcal{C}_{#1}^{'}}\xspace}                    
\def\Ceff#1   {\ensuremath{\mathcal{C}_{#1}^{\mathrm{(eff)}}}\xspace}        
\def\Cpeff#1  {\ensuremath{\mathcal{C}_{#1}^{'\mathrm{(eff)}}}\xspace}       
\def\Ope#1    {\ensuremath{\mathcal{O}_{#1}}\xspace}                       
\def\Opep#1   {\ensuremath{\mathcal{O}_{#1}^{'}}\xspace}                    

\def\ycp        {\ensuremath{y_{\CP}}\xspace}

\newcommand{\ket}[1]{\ensuremath{|#1\rangle}}              

\newcommand{\tev}{\ifthenelse{\boolean{inbibliography}}{\ensuremath{~T\kern -0.05em eV}}{\ensuremath{\mathrm{\,Te\kern -0.1em V}}}\xspace}
\newcommand{\gev}{\ensuremath{\mathrm{\,Ge\kern -0.1em V}}\xspace}
\newcommand{\mev}{\ensuremath{\mathrm{\,Me\kern -0.1em V}}\xspace}
\newcommand{\kev}{\ensuremath{\mathrm{\,ke\kern -0.1em V}}\xspace}
\newcommand{\ev}{\ensuremath{\mathrm{\,e\kern -0.1em V}}\xspace}
\newcommand{\gevc}{\ensuremath{{\mathrm{\,Ge\kern -0.1em V\!/}c}}\xspace}
\newcommand{\mevc}{\ensuremath{{\mathrm{\,Me\kern -0.1em V\!/}c}}\xspace}
\newcommand{\gevcc}{\ensuremath{{\mathrm{\,Ge\kern -0.1em V\!/}c^2}}\xspace}
\newcommand{\gevgevcccc}{\ensuremath{{\mathrm{\,Ge\kern -0.1em V^2\!/}c^4}}\xspace}
\newcommand{\mevcc}{\ensuremath{{\mathrm{\,Me\kern -0.1em V\!/}c^2}}\xspace}

\def\invpb {\ensuremath{\mbox{\,pb}^{-1}}\xspace}

\def\invfb   {\ensuremath{\mbox{\,fb}^{-1}}\xspace}

\def\ps   {\ensuremath{{\mathrm{ \,ps}}}\xspace}

\def\invps{\ensuremath{{\mathrm{ \,ps^{-1}}}}\xspace}

\newcommand{\stat}{\ensuremath{\mathrm{\,(stat)}}\xspace}
\newcommand{\syst}{\ensuremath{\mathrm{\,(syst)}}\xspace}

\newcommand{\chisq}{\ensuremath{\chi^2}\xspace}

\newcommand{\chisqip}{\ensuremath{\chi^2_{\text{IP}}}\xspace}

\def\gsim{{~\raise.15em\hbox{$>$}\kern-.85em
          \lower.35em\hbox{$\sim$}~}\xspace}
\def\lsim{{~\raise.15em\hbox{$<$}\kern-.85em
          \lower.35em\hbox{$\sim$}~}\xspace}

\def\tell1  {TELL1\xspace}
\def\ukl1   {UKL1\xspace}

\newcommand{\ie}{\mbox{\itshape i.e.}\xspace}

%% file: title-LHCb-PAPER.tex

\begin{titlepage}
\pagenumbering{roman}

\vspace*{-1.5cm}
\centerline{\large EUROPEAN ORGANIZATION FOR NUCLEAR RESEARCH (CERN)}
\vspace*{1.5cm}
\noindent
\begin{tabular*}{\linewidth}{lc@{\extracolsep{\fill}}r@{\extracolsep{0pt}}}
\ifthenelse{\boolean{pdflatex}}
{\vspace*{-1.5cm}\mbox{\!\!\!\includegraphics[width=.14\textwidth]{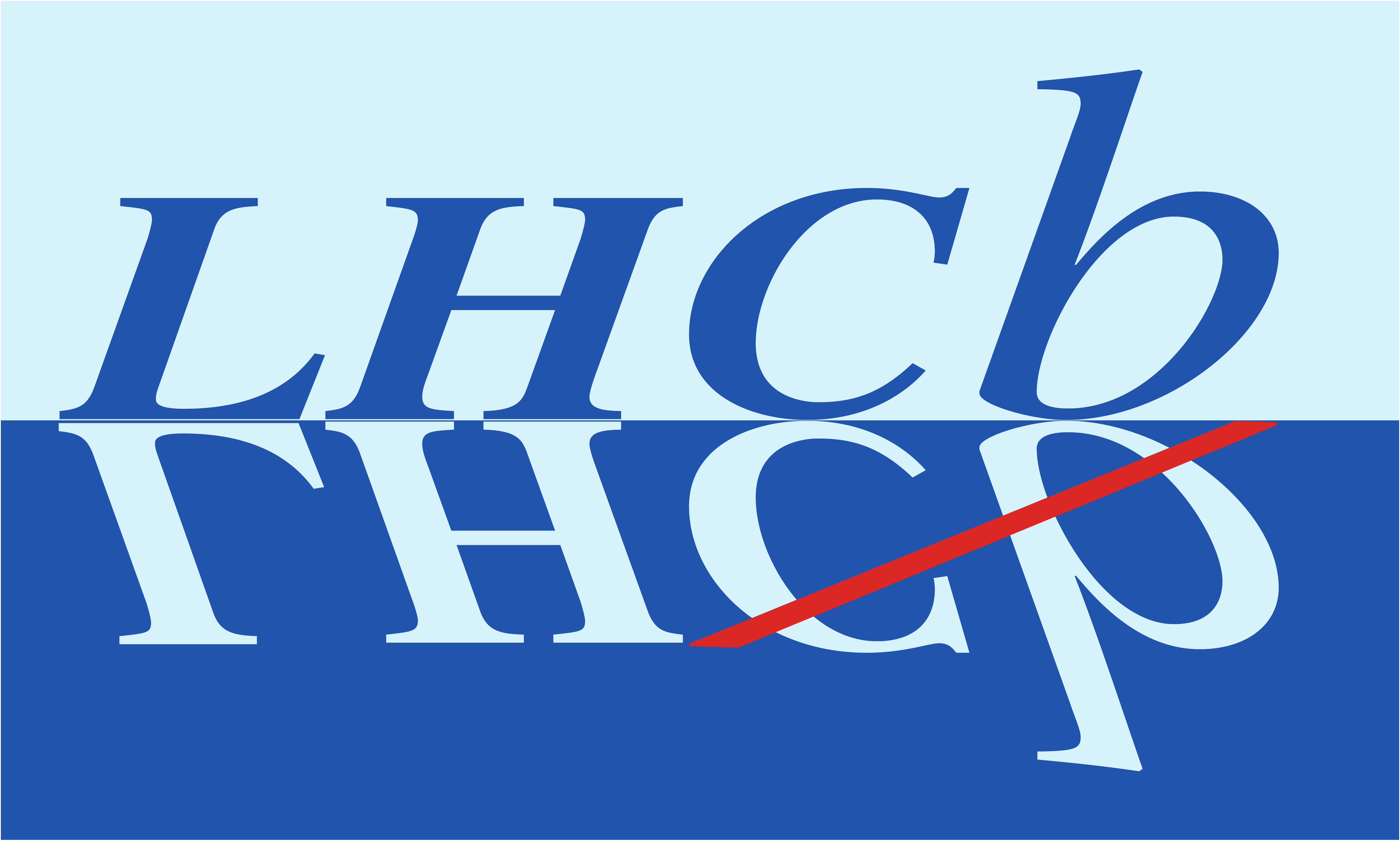}} & &}%
{\vspace*{-1.2cm}\mbox{\!\!\!\includegraphics[width=.12\textwidth]{figs/lhcb-logo.eps}} & &}%
\\
 & & \mycernpapernumber \\  
 & & \mylhcbpapernumber \\  
 & & \mydate \\ 
 & & \mypaperversion\\
\end{tabular*}

\vspace*{4.0cm}

{\normalfont\bfseries\boldmath\huge
\begin{center}
  \papertitle 
\end{center}
}

\vspace*{2.0cm}

\begin{center}
\paperauthors\footnote{Authors are listed at the end of the Letter.}
\end{center}

\vspace{\fill}

\begin{abstract}
\input{abstract}
\end{abstract}

\vspace*{2.0cm}

\begin{center}
  Published in Phys.~Rev.~Lett. 122 (2019) 011802
\end{center}

\vspace{\fill}

{\footnotesize 
\centerline{\copyright~\papercopyright. \href{\paperlicenceurl}{\paperlicence}.}}
\vspace*{2mm}

\end{titlepage}


\newpage
\setcounter{page}{2}
\mbox{~}
%
%
%
%

\cleardoublepage

%% file: abstract.tex
\noindent A measurement of the charm-mixing parameter \ycp using \dzkk, \dzpipi, and \dzkpi decays is reported. The \Dz mesons are required to originate from semimuonic decays of \Bub and \Bzb mesons. These decays are partially reconstructed in a data set of proton-proton collisions at center-of-mass energies of 7 and 8\tev collected with the LHCb experiment and corresponding to an integrated luminosity of 3\invfb. The \ycp parameter is measured to be $(\resultycppc \pm \errstatycppc\stat \pm \errsystycppc\syst)\%$, in agreement with, and as precise as, the current world-average value.

%% file: body.tex
Neutral charm mesons can change their flavor and turn into antimesons, and vice versa, before they decay. This phenomenon, known as flavor oscillation or \Dz--\Dzb mixing, occurs because the eigenstates of the Hamiltonian governing the time evolution of the neutral \D system are superpositions of the flavor eigenstates, ${\ket{D_{1,2}} = p\ket{\Dz} \pm q \ket{\Dzb}}$, where $p$ and $q$ are complex parameters satisfying $|p|^2+|q|^2=1$. In the limit of charge-parity (\CP) symmetry, $q$ equals $p$ and the oscillations are characterized by only two dimensionless parameters, $x \equiv  (m_1 - m_2)/\Gamma$ and $y \equiv (\Gamma_1 - \Gamma_2)/ 2\Gamma$, where $m_{1(2)}$ and $\Gamma_{1(2)}$ are the mass and decay width of the \CP-even (odd) eigenstate $D_{1(2)}$, respectively, and $\Gamma \equiv (\Gamma_1 + \Gamma_2)/2$ is the average decay width~\cite{PDG2018}. The values of $x$ and $y$ are of the order of 1\% or smaller \cite{HFLAV16}. In the presence of \CP violation, the mixing rates for mesons produced as \Dz and \Dzb differ, further enriching the phenomenology.

Because of \Dz--\Dzb mixing, the \emph{effective} decay width $\Gamma_{\CP+}$ of decays to \CP-even final states, such as $h^+h^-$ ($h=K$, $\pi$), differs from the average width $\Gamma$. The latter can be measured in decays that involve an equal mixture of \CP-even and \CP-odd  states, such as \dzkpi.\footnote{Throughout this Letter, the inclusion of the charge-conjugate decay mode is implied unless otherwise stated.} The quantity
\begin{equation}
\ycp\equiv\frac{\Gamma_{\CP+}}{\Gamma}-1
\end{equation}
is equal to the mixing parameter $y$ if \CP symmetry is conserved. Otherwise, it is related to $x$, $y$, $|q/p|$, and $\phi\equiv\arg(q\overline{\mathcal{A}}/p\mathcal{A})$, as 
${2\ycp\approx\left(\left|q/p\right|+\left|p/q\right|\right)y\cos\phi - \left(\left|q/p\right|-\left|p/q\right|\right) x \sin\phi}$, where $\mathcal{A}$ ($\overline{\mathcal{A}}$) is the \Dz (\Dzb) decay amplitude~\cite{Du:1986ai,Bergmann:2000id}. The approximation holds for decays, such as \dzhh, that can be described by a single amplitude. Neglecting the $\mathcal{O}(10^{-3})$ difference between the phases of the \dzkk and \dzpipi decay amplitudes, $\phi$~is universal and \ycp is independent of the $h^+h^-$ final state.

The current world average value of \ycp, $(0.84\pm 0.16)\%$~\cite{HFLAV16}, is dominated by measurements at the \B factories~\cite{Lees:2012qh,Staric:2015sta} and is consistent with the value of $y$, $(0.62 \pm 0.07)\%$~\cite{HFLAV16}. The only measurement of \ycp at a hadron collider, $(0.55\pm0.63\stat\pm0.41\syst)\%$, has been made by the LHCb collaboration using a sample of proton-proton collisions corresponding to an integrated luminosity of $29\invpb$~\cite{LHCb-PAPER-2011-032}. Improving the precision of both \ycp and $y$ might lead to evidence of \CP violation in \Dz--\Dzb mixing if they differ significantly. This would offer sensitivity to a broad class of non-standard-model processes that could contribute to the mixing amplitude by increasing the oscillation rate and/or introducing \CP-violation effects that are highly suppressed in the standard model~\cite{Blaylock:1995ay,Bianco:2003vb,Grossman:2006jg,Petrov:2006nc,Golowich:2007ka,Ciuchini:2007cw}. Searches for \CP violation in the up-quark sector are also complementary to those performed with beauty and strange mesons, thus providing a unique opportunity to make progress in the understanding of the mechanisms responsible for the observed asymmetry between matter and antimatter in the Universe~\cite{Sakharov:1967dj,Huet:1994jb}.

In this Letter, a measurement of \ycp using \dzkk, \dzpipi, and \dzkpi decays is reported. The \Dz mesons are required to originate from semimuonic decays of \Bub or \Bzb mesons, collectively referred to as \bdzmux. The difference between the widths of \Dz decays to \CP-even and \CP-mixed final states,
\begin{equation}
\deltag \equiv \Gamma_{\CP+} - \Gamma\,,
\end{equation}
is measured from a fit to the ratio between \dzkk (or \dzpipi) and \dzkpi signal yields as a function of the \Dz decay time. The parameter \ycp is then calculated from the measured value of \deltag and the precisely known value of $\Gamma$~\cite{PDG2018} as ${\ycp = \deltag / \Gamma}$. The \Dz decay time is defined as $t = (m\,\vec{L} \cdot \vec{p}\,)/|\vec{p}\,|^{2}$, where $m$ is the known value of the \Dz mass~\cite{PDG2018}, $\vec{L}$ is the vector connecting the \Bb and the \Dz decay vertices, and $\vec{p}$ is the momentum of the \Dz meson. The selection efficiency as a function of the \Dz decay time (decay-time acceptance) is very similar for \dzhh and \dzkpi decays. However, since the average opening angle of a two-body decay in the laboratory frame depends on the masses of its decay products, differences of the order of a few percent are present and are corrected for in the analysis. The correction is evaluated using simulation and validated using control samples of data, which also include \dkpipi and \dkkpi decays with \Dp decays originating from semimuonic \Bb decays (referred to as \bdmux). To avoid potential experimenter's bias, the measured value of \ycp remained unknown during the development of the analysis and was examined only after the analysis procedure and the evaluation of the systematic uncertainties were finalized.

Semileptonic decays of \Bb mesons are partially reconstructed in a data set collected with the LHCb experiment in \proton\proton collisions at center-of-mass energies of 7 and $8\tev$ and corresponding to an integrated luminosity of $3\invfb$. The \lhcb detector is a single-arm forward spectrometer equipped with precise charged-particle vertexing and tracking detectors, hadron-identification detectors, calorimeters, and muon detectors, optimized for the study of bottom- and charm-hadron decays~\cite{Alves:2008zz,LHCb-DP-2014-002}. Simulation~\cite{LHCb-PROC-2010-056,LHCb-PROC-2011-006,LHCb-DP-2018-004} is used to model all relevant sources of decays, correct the data for the decay-time acceptance, study the decay-time resolution, and evaluate systematic uncertainties on the measurement.

The online event selection is performed by a trigger that consists of a hardware stage, based on information from the calorimeter and muon systems, followed by a two-level software stage, which applies a full event reconstruction~\cite{LHCb-DP-2012-004}. To select semimuonic \Bb decays, the hardware trigger requires a muon candidate with transverse momentum exceeding $1.5$ to $1.8\gevc$, depending on the data-taking period. In the first level of the software trigger, the selected muon is required to be displaced from any \proton\proton interaction point. These requirements do not bias the decay time of the \D candidate. In the second level of the software trigger, the muon candidate is associated with one, two, or three charged particles, all displaced from the same \proton\proton interaction point. This association can bias the decay time, favoring shorter \D flight distances, as the muon and the \D decay products satisfying the trigger criteria must be consistent with originating from a common displaced vertex.

In the offline reconstruction, the muon candidate is combined with charged particles, forming the \D-meson candidate and identified to be either kaons or pions, according to the topology and kinematics of \bdzmux and \bdmux decays. The requirements to select \bdzmux decays are inherited from the analysis reported in Ref.~\cite{LHCb-PAPER-2014-013}; those for \bdmux decays are taken from Ref.~\cite{LHCb-PAPER-2017-004}. In these selections, the \D decay products are requested to be displaced from the \proton\proton interaction point with respect to which they have the smallest \chisqip, by imposing $\chisqip>9$. The \chisqip\ is defined as the difference in the vertex-fit \chisq of a given interaction point reconstructed with and without the particle being considered. These requirements are particularly relevant for the measurement of \ycp as they bias the \D decay-time distribution, being more efficient for decays with a larger flight distance. The following additional requirements, not used in Refs.~\cite{LHCb-PAPER-2014-013,LHCb-PAPER-2017-004}, are applied. The $\D\mu$ invariant mass, $m(\D\mu)$, must not exceed $5.2\gevcc$, to suppress genuine charm decays accidentally combined with unrelated muon candidates. The mass of the \D candidate must be in the range $1.825$--$1.920\gevcc$. Its decay time must be larger than $0.15\ps$ to minimize a bias observed in simulation at $t\approx0$ due to the reconstruction of the \Bb vertex. A requirement on the component of the \D momentum transverse to the \Bb flight direction is applied as a function of the corrected \Bb mass to suppress decays of \bquark hadrons into final states with a pair of charm hadrons, of which one decays semileptonically, and background from semitauonic decays $\Bb \to \D\tau^-\neutb X$, with $\tau^-\to\mu^-\neumb\neut$. The corrected \Bb mass is determined from the $\D\mu$ invariant mass as $\sqrt{m^2(\D\mu)+p_\perp^2(\D\mu)}+p_\perp(\D\mu)$, using the momentum of the $\D\mu$ system transverse to the \Bb flight direction, $p_\perp(\D\mu)$, to partially compensate for the momentum of the unreconstructed decay products. After the selection, these background contributions total to at most $1.5\%$ of the signal yield. A contamination of about 1\% of \D decays produced directly in the \proton\proton collision (prompt \D) is also estimated to be present in the selected sample. All these background decays are checked to have negligible impact on the measurement of \ycp.

\begin{figure}[t]
\centering
\includegraphics[width=0.55\textwidth]{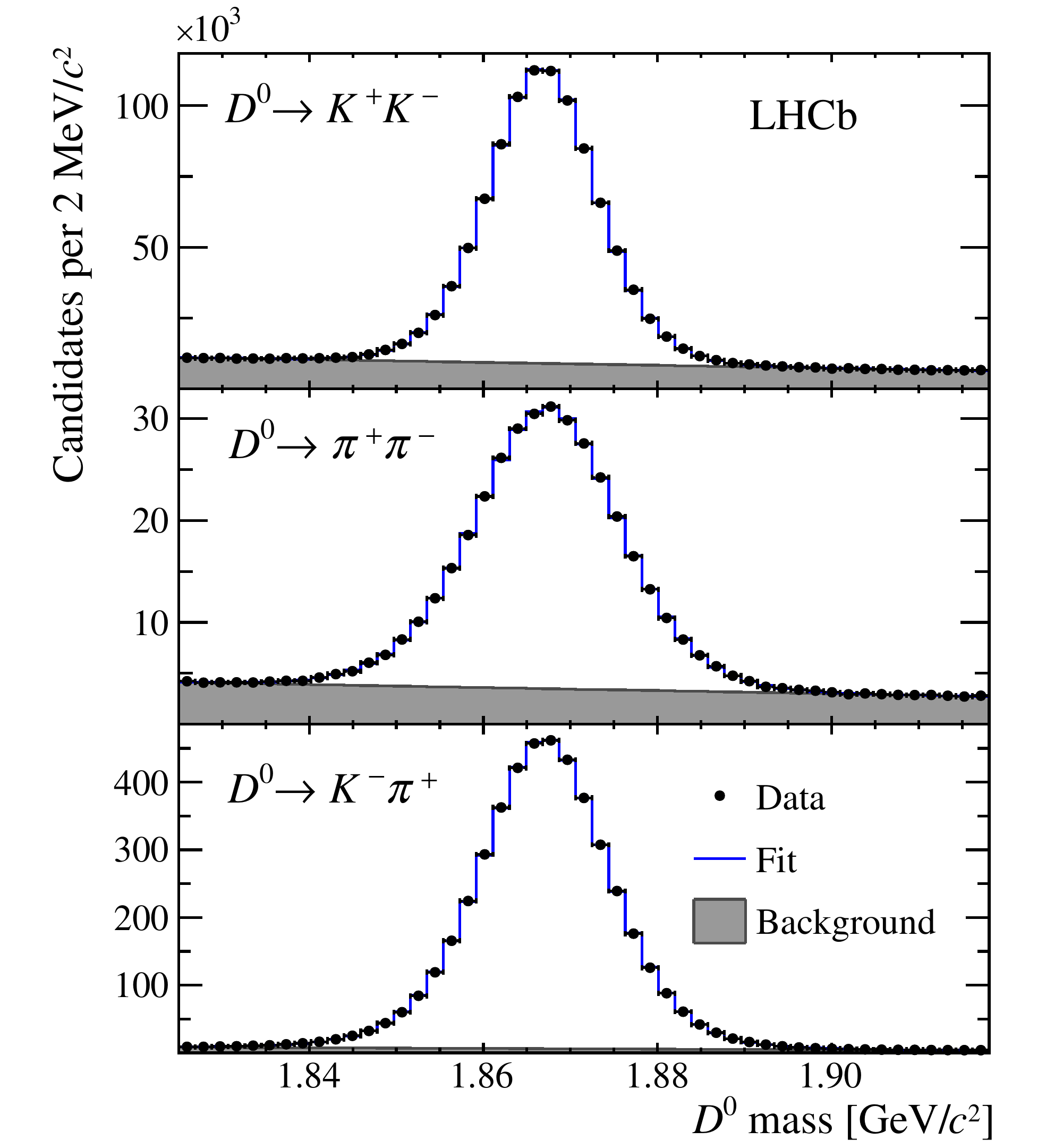}
\caption{Distribution of \Dz mass for candidates passing the selection with fit projections overlaid: (top) \dzkk decays, (center) \dzpipi decays, and (bottom) \dzkpi decays.\label{fig:D_Mass}}
\end{figure}

Figure~\ref{fig:D_Mass} shows the \Dz mass distributions of the selected candidates. Prominent signal peaks at the known \Dz mass values are visible on top of a smooth background made of random combinations of charged particles faking a \Dz candidate. The small contamination of prompt \Dz decays is included in the signal peak. Binned $\chi^2$ fits to the mass distributions determine the signal yields reported in Table~\ref{tab:signal_yields}, together with the yields of the control samples of \Dp decays. The fits use a probability density function (pdf) consisting of a Johnson $S_U$ distribution~\cite{johnson} (or the sum of a Johnson $S_U$ and a Gaussian distribution in the case of \dzkpi and \dkpipi decays) to describe the asymmetric shape of the signal peak, and a linear distribution to describe the background.  

\begin{table}[t]
\caption{Signal yields of the selected candidates.}\label{tab:signal_yields}
\centering
\begin{tabular}{lr}
\toprule
Decay & Signal yield $[10^3]$ \\
\midrule
\dzkk   &  $878.2 \pm 1.2$ \\
\dzpipi &  $311.6 \pm 0.9$ \\
\dzkpi  & $4579.5 \pm 3.2$ \\
\dkpipi & $2260.2 \pm 1.9$ \\
\dkkpi  &   $98.0 \pm 0.3$ \\
\bottomrule
\end{tabular}
\end{table}

\begin{figure}[b]
\centering
\includegraphics[width=0.65\textwidth]{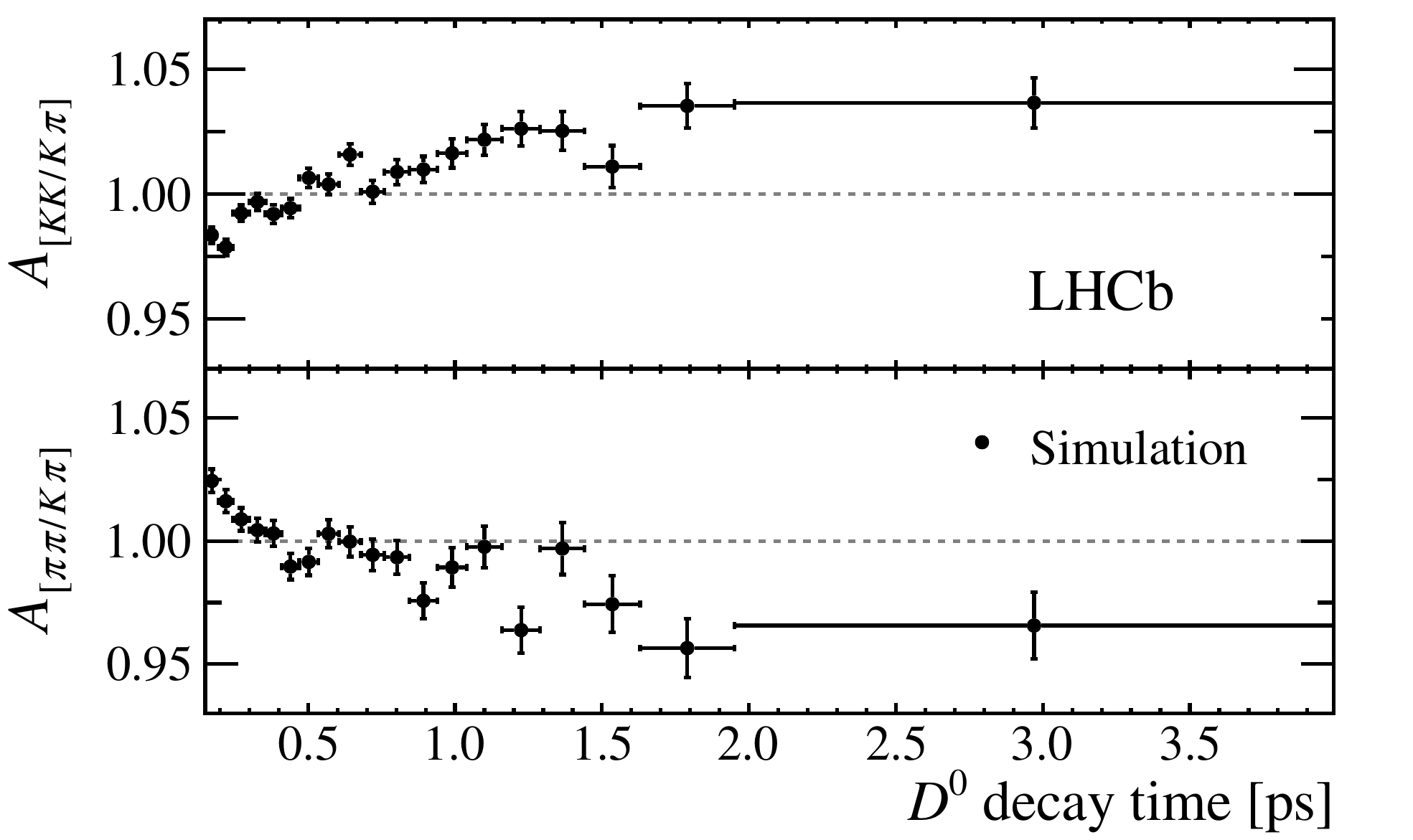} 
\caption{Ratio of decay-time acceptances from simulation for (top) \dzkk over \dzkpi decays and (bottom) \dzpipi over \dzkpi decays.\label{fig:acc_ratio}}
\end{figure}

The sample is split into 19 disjoint subsets (bins) of \D decay time spanning the range $0.15$--$4\ps$. The signal yields are determined in each decay-time bin with fits to the \D mass distribution using the same pdf as described above. In these fits all signal-shape parameters are fixed to the values from the decay-time-integrated fits, with the exception of the mean and width of the Johnson function. The ratio between \dzkk (or \dzpipi) and  \dzkpi signal yields as a function of decay time is fitted to determine the value of \deltag. The fit minimizes a $\chi^2$ function where the signal-yield ratio in a decay-time bin is described by the ratio of the integrals of two decreasing exponential functions, one for the numerator with exponent $\Gamma_{\CP+} = \deltag + \Gamma$, and the other for the denominator with exponent $\Gamma$. The value of $\Gamma$ is fixed to its world average of $2.4384\invps$~\cite{PDG2018}, while \deltag and a decay-time-independent normalization factor of the ratio are free to vary in the fit. It should be noted that $\Gamma$ can be fixed to any arbitrary value, since the distribution of the ratio is only sensitive to \deltag. In the fit, the signal-yield ratio is corrected in each decay-time bin by a factor calculated as the ratio of the decay-time acceptances of the decays in the numerator and the denominator. This correction is determined from simulation and shows up to $6\%$ variations around unity as a function of \Dz decay time (Figure~\ref{fig:acc_ratio}). The correction is similar in magnitude, but with an opposite trend as a function of $t$, for the determination of \deltag with \dzkk and \dzpipi decays.

Several null tests are performed on data to prove that the estimates of the signal yields are unbiased, and that the corrections from simulation are reliable. The tests use samples of (i)~\dkkpi and \dkpipi decays, (ii)~\dkpipi decays, (iii)~\dzkpi decays, and (iv)~\dzkk decays. In test (i), the width difference is measured by fitting the yield ratio of \dkkpi to \dkpipi decays. The corrections for the ratio of decay-time acceptances are similar to those in the \ycp measurement. In  tests (ii)--(iv), the selected data are split randomly into two independent sets: one is used as the denominator sample, and the other, featuring a tighter requirement of $\chisqip>60$ for the \D decay products, is used as the numerator sample. The threshold on $\chisqip$ is chosen such that the ratio of decay-time acceptances deviates from a constant by up to 40\%, \ie, almost an order of magnitude larger variation than that present in the \ycp measurement. In all tests, the measured decay-width difference is consistent with zero, with fit $p$ values ranging from 8\% to 84\%. The two most precise tests, (ii) and (iii), correspond to a validation of the measurement of \ycp with an uncertainty of 0.14\%, which includes the limited knowledge of the decay-time acceptance correction. Another test (v) consists in measuring the decay-width difference of \Dp and \Dz mesons, using the largest-yield samples of \dkpipi and \dzkpi decays. In this measurement, the ratio of the decay-time acceptances presents variations up to about 10\%. However, the decays considered in the numerator and the denominator have sufficiently different topologies that potential biases on the measurement of the width difference are not suppressed in the ratio at the same level as in the \ycp measurement. In addition, the very different lifetimes between \Dp and \Dz mesons lead to a signal-yield ratio spanning over a very broad interval, with a maximum approximately 25 times larger than its minimum. The ratio of \Dp to \Dz lifetimes is determined to be $2.5141 \pm 0.0082$, where the uncertainty is only statistical, in agreement with the known value of $2.536 \pm 0.019$~\cite{PDG2018}. Biases that scale with \deltag are excluded by this test within a relative precision of about $1\%$. In summary, the five tests yield results consistent with the expectations with a $\chi^2$ of $5.5$, which corresponds to a $p$ value of $36\%$. The tests demonstrate that the acceptance effects needed for the measurement of \ycp are understood within the precision provided by the limited size of the simulated samples. The tests also confirm that background originating from prompt \D decays, from \bquark-hadron decays to double-charm final states, and from semitauonic \Bb decays can be neglected. They contaminate all samples considered in the tests with fractions similar to those estimated in the \ycp measurement.

\begin{figure}[t]
\centering
\includegraphics[width=0.65\textwidth]{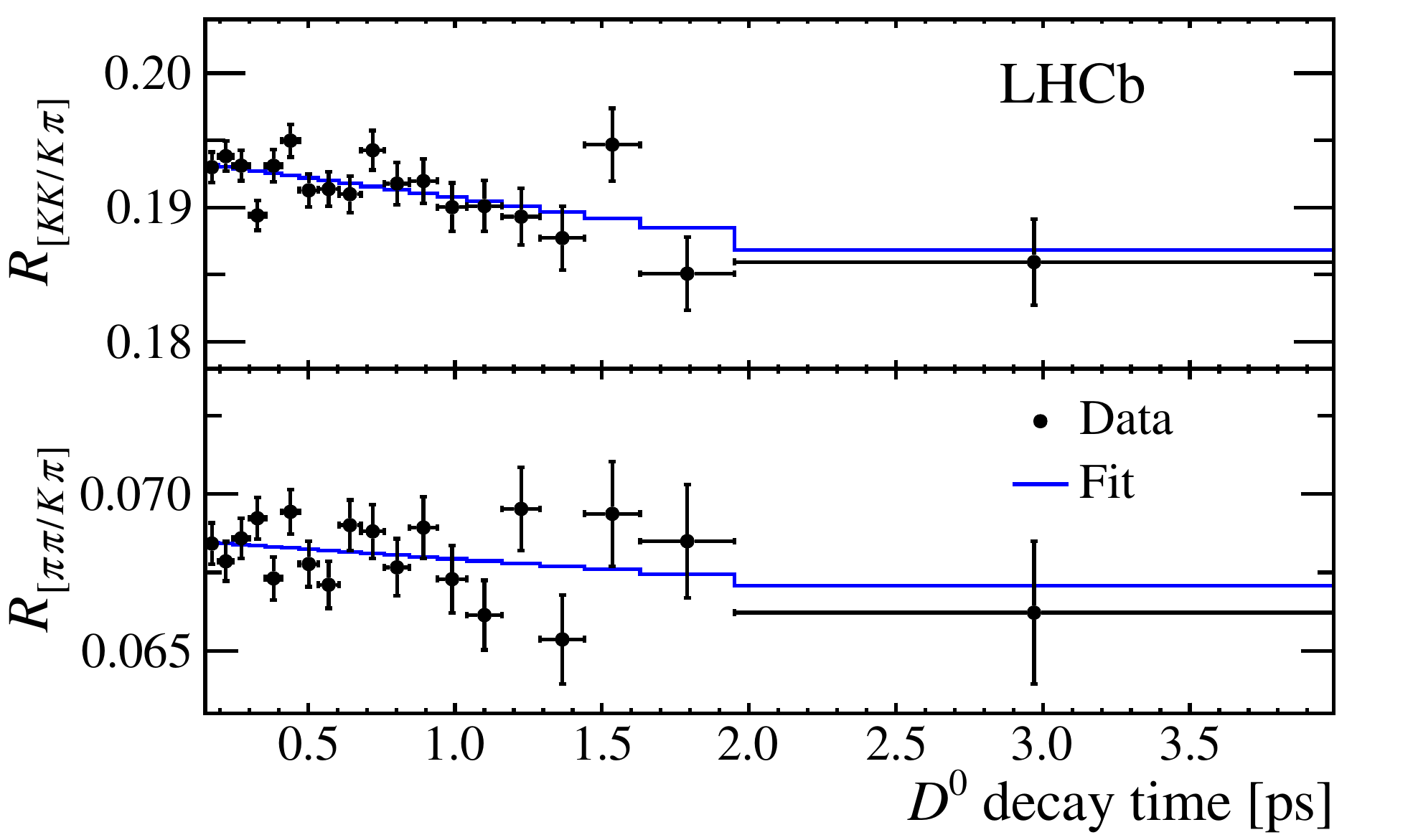}
\caption{Acceptance-corrected signal-yield ratio of (top) \dzkk over \dzkpi decays and (bottom) \dzpipi over \dzkpi decays as a function of \Dz decay time, with fit projection overlaid.\label{fig:fits}}
\end{figure}

Figure~\ref{fig:fits} shows the acceptance-corrected signal-yield ratio measured for the \dzkk and \dzpipi decays with respect to \dzkpi decays, with fit projections overlaid. The obtained values of \deltag and \ycp are reported in Table~\ref{tab:results}. The use of a common reference sample (\dzkpi) does not introduce any significant correlation between the statistical uncertainties of the \dzkk and \dzpipi measurements.

\begin{table}[t]
\caption{Measured values of \deltag and \ycp. The first uncertainty is statistical, the second is systematic.\label{tab:results}}
\centering
\begin{tabular}{lcc}
\toprule
Decay & \deltag $[\invps]$ & \ycp $[\%]$\\
\midrule
\dzkk & $\resultdgk\pm\errstatonlydgk\pm\errsystdgk$ & $\resultycpkpc\pm\errstatonlyycpkpc\pm\errsystycpkpc$ \\
\dzpipi & $\resultdgpi\pm\errstatonlydgpi\pm\errsystdgpi$ & $\resultycppipc\pm\errstatonlyycppipc\pm\errsystycppipc$ \\ 
\bottomrule
\end{tabular}
\end{table}

Systematic uncertainties of $\errsystdgk\invps$ ($\errsystdgpi\invps$) on \deltag, and therefore $\errsystycpkpc\%$ ($\errsystycppipc\%$) on \ycp, are assigned for the measurement done with \dzkk (\dzpipi) decays. The correlation between the systematic uncertainties is 5\%. They are dominated by the knowledge of the correction for the ratio of decay-time acceptances, which is limited by the finite size of the simulated samples. This yields systematic uncertainties of $0.0026\invps$ ($0.0037\invps$) on \deltag and $0.11\%$ ($0.15\%$) on \ycp, which are uncorrelated between the \dzkk and \dzpipi measurements. Other systematic uncertainties, contributing less, are associated with the assumed decay model and composition of the simulated samples of semileptonic \Bb decays ($0.0006\invps$ on \deltag, $0.02\%$ on \ycp), possible biases introduced by the fit method as determined in large ensembles of pseudoexperiments ($0.0004\invps$ on \deltag, $0.02\%$ on \ycp), and the neglected $0.12\ps$ decay-time resolution ($0.0003\invps$ on \deltag, $0.01\%$ on \ycp). These systematic uncertainties are fully correlated between the measurements  with \dzkk and \dzpipi decays. Asymmetric production of \Dz and \Dzb mesons from semileptonic \Bub and \Bzb decays produce biases on \ycp that are smaller than $10^{-5}$. Uncertainties on the measured decay-length arising from relative misalignments of subdetectors and the uncertainty of the input value of $\Gamma$, $2.4384\pm0.0089\invps$~\cite{PDG2018}, which is used to determine \ycp from \deltag, have negligible contributions. Finally, consistency checks based on repeating the \ycp measurement on independent subsamples chosen according to data-taking periods, trigger-selection criteria and interaction-point multiplicity all yield compatible results within statistical fluctuations.

In summary, the charm mixing-parameter \ycp is measured using \dzkk, \dzpipi and \dzkpi decays originating from semileptonic \Bub and \Bzb decays produced in \proton\proton collision data collected with the LHCb experiment at center-of-mass energies of 7 and 8\tev, and corresponding to an integrated luminosity of 3\invfb. The results from \dzkk, $\ycp =(\resultycpkpc\pm\errstatonlyycpkpc\stat\pm\errsystycpkpc\syst)\%$, and \dzpipi decays, $\ycp = (\resultycppipc\pm\errstatonlyycppipc\stat\pm\errsystycppipc\syst)\%$, are consistent with each other and with determinations from other experiments~\cite{HFLAV16}. The value of \ycp measured in the \dzkk mode is the most precise to date from a single experiment. The two measurements are combined and yield $\ycp = (\resultycppc \pm \errstatycppc\stat \pm \errsystycppc\syst)\%$, which is consistent with and as precise as the current world average value, $(0.84 \pm 0.16)\%$~\cite{HFLAV16}. The result is also consistent with the known value of the mixing parameter $y$, $(0.62 \pm 0.07)\%$~\cite{HFLAV16}, showing no evidence for \CP violation in \Dz--\Dzb mixing. As larger data samples are accumulated by LHCb, the dominant systematic uncertainty due to finite simulation samples will also be reduced, giving good prospects for further reduction in the uncertainty of \ycp.

%% file: acknowledgements.tex
\section*{Acknowledgements}
%
%
\noindent We express our gratitude to our colleagues in the CERN
accelerator departments for the excellent performance of the LHC. We
thank the technical and administrative staff at the LHCb
institutes.
We acknowledge support from CERN and from the national agencies:
CAPES, CNPq, FAPERJ and FINEP (Brazil); 
MOST and NSFC (China); 
CNRS/IN2P3 (France); 
BMBF, DFG and MPG (Germany); 
INFN (Italy); 
NWO (Netherlands); 
MNiSW and NCN (Poland); 
MEN/IFA (Romania); 
MSHE (Russia); 
MinECo (Spain); 
SNSF and SER (Switzerland); 
NASU (Ukraine); 
STFC (United Kingdom); 
NSF (USA).
We acknowledge the computing resources that are provided by CERN, IN2P3
(France), KIT and DESY (Germany), INFN (Italy), SURF (Netherlands),
PIC (Spain), GridPP (United Kingdom), RRCKI and Yandex
LLC (Russia), CSCS (Switzerland), IFIN-HH (Romania), CBPF (Brazil),
PL-GRID (Poland) and OSC (USA).
We are indebted to the communities behind the multiple open-source
software packages on which we depend.
Individual groups or members have received support from
AvH Foundation (Germany);
EPLANET, Marie Sk\l{}odowska-Curie Actions and ERC (European Union);
ANR, Labex P2IO and OCEVU, and R\'{e}gion Auvergne-Rh\^{o}ne-Alpes (France);
Key Research Program of Frontier Sciences of CAS, CAS PIFI, and the Thousand Talents Program (China);
RFBR, RSF and Yandex LLC (Russia);
GVA, XuntaGal and GENCAT (Spain);
the Royal Society
and the Leverhulme Trust (United Kingdom);
Laboratory Directed Research and Development program of LANL (USA).

%% file: LHCb_Authorship_flat_28-Aug-2018/LHCb_Authorship_28-Aug-2018.tex
\centerline
{\large\bf LHCb Collaboration}
\begin
{flushleft}
\small
R.~Aaij$^{29}$,
C.~Abell{\'a}n~Beteta$^{46}$,
B.~Adeva$^{43}$,
M.~Adinolfi$^{50}$,
C.A.~Aidala$^{77}$,
Z.~Ajaltouni$^{7}$,
S.~Akar$^{61}$,
P.~Albicocco$^{20}$,
J.~Albrecht$^{12}$,
F.~Alessio$^{44}$
% ...

{\footnotesize \it

% ...
$ ^{7}$Universit{\'e} Clermont Auvergne, CNRS/IN2P3, LPC, Clermont-Ferrand, France\\
% ...
$ ^{12}$Fakult{\"a}t Physik, Technische Universit{\"a}t Dortmund, Dortmund, Germany\\
% ...
$ ^{20}$INFN Laboratori Nazionali di Frascati, Frascati, Italy\\
% ...
$ ^{29}$Nikhef National Institute for Subatomic Physics, Amsterdam, Netherlands\\
% ...
$ ^{43}$Instituto Galego de F{\'\i}sica de Altas Enerx{\'\i}as (IGFAE), Universidade de Santiago de Compostela, Santiago de Compostela, Spain\\
$ ^{44}$European Organization for Nuclear Research (CERN), Geneva, Switzerland\\
% ...
$ ^{46}$Physik-Institut, Universit{\"a}t Z{\"u}rich, Z{\"u}rich, Switzerland\\
% ...
$ ^{50}$H.H. Wills Physics Laboratory, University of Bristol, Bristol, United Kingdom\\
% ...
$ ^{61}$University of Cincinnati, Cincinnati, OH, United States\\
% ...
$ ^{63}$Syracuse University, Syracuse, NY, United States\\
% ...
$ ^{77}$University of Michigan, Ann Arbor, United States, associated to $^{63}$\\
% ...
}
\end{flushleft}

%% file: main.bbl
\ifx\mcitethebibliography\mciteundefinedmacro
\PackageError{LHCb.bst}{mciteplus.sty has not been loaded}
{This bibstyle requires the use of the mciteplus package.}\fi
\providecommand{\href}[2]{#2}
\begin{mcitethebibliography}{10}
\mciteSetBstSublistMode{n}
\mciteSetBstMaxWidthForm{subitem}{\alph{mcitesubitemcount})}
\mciteSetBstSublistLabelBeginEnd{\mcitemaxwidthsubitemform\space}
{\relax}{\relax}

\bibitem{PDG2018}
Particle Data Group, M.~Tanabashi {\em et~al.},
  \ifthenelse{\boolean{articletitles}}{\emph{{\href{http://pdg.lbl.gov/}{Review
  of particle physics}}}, }{}Phys.\ Rev.\  \textbf{D98} (2018) 030001\relax
\mciteBstWouldAddEndPuncttrue
\mciteSetBstMidEndSepPunct{\mcitedefaultmidpunct}
{\mcitedefaultendpunct}{\mcitedefaultseppunct}\relax
\EndOfBibitem
\bibitem{HFLAV16}
Heavy Flavor Averaging Group, Y.~Amhis {\em et~al.},
  \ifthenelse{\boolean{articletitles}}{\emph{{Averages of $b$-hadron,
  $c$-hadron, and $\tau$-lepton properties as of summer 2016}},
  }{}\href{https://doi.org/10.1140/epjc/s10052-017-5058-4}{Eur.\ Phys.\ J.\
  \textbf{C77} (2017) 895},
  \href{http://arxiv.org/abs/1612.07233}{{\normalfont\ttfamily
  arXiv:1612.07233}}, {updated results and plots available at
  \href{https://hflav.web.cern.ch}{{\texttt{https://hflav.web.cern.ch}}}}\relax
\mciteBstWouldAddEndPuncttrue
\mciteSetBstMidEndSepPunct{\mcitedefaultmidpunct}
{\mcitedefaultendpunct}{\mcitedefaultseppunct}\relax
\EndOfBibitem
\bibitem{Du:1986ai}
D.~S. Du, \ifthenelse{\boolean{articletitles}}{\emph{{Searching for possible
  large \CP-violation effects in neutral-charm-meson decays}},
  }{}\href{https://doi.org/10.1103/PhysRevD.34.3428}{Phys.\ Rev.\  \textbf{D34}
  (1986) 3428}\relax
\mciteBstWouldAddEndPuncttrue
\mciteSetBstMidEndSepPunct{\mcitedefaultmidpunct}
{\mcitedefaultendpunct}{\mcitedefaultseppunct}\relax
\EndOfBibitem
\bibitem{Bergmann:2000id}
S.~Bergmann {\em et~al.}, \ifthenelse{\boolean{articletitles}}{\emph{{Lessons
  from CLEO and FOCUS measurements of \Dz--\Dzb mixing parameters}},
  }{}\href{https://doi.org/10.1016/S0370-2693(00)00772-3}{Phys.\ Lett.\
  \textbf{B486} (2000) 418},
  \href{http://arxiv.org/abs/hep-ph/0005181}{{\normalfont\ttfamily
  arXiv:hep-ph/0005181}}\relax
\mciteBstWouldAddEndPuncttrue
\mciteSetBstMidEndSepPunct{\mcitedefaultmidpunct}
{\mcitedefaultendpunct}{\mcitedefaultseppunct}\relax
\EndOfBibitem
\bibitem{Lees:2012qh}
BaBar collaboration, J.~P. Lees {\em et~al.},
  \ifthenelse{\boolean{articletitles}}{\emph{{Measurement of \Dz--\Dzb mixing
  and \CP violation in two-body \Dz decays}},
  }{}\href{https://doi.org/10.1103/PhysRevD.87.012004}{Phys.\ Rev.\
  \textbf{D87} (2013) 012004},
  \href{http://arxiv.org/abs/1209.3896}{{\normalfont\ttfamily
  arXiv:1209.3896}}\relax
\mciteBstWouldAddEndPuncttrue
\mciteSetBstMidEndSepPunct{\mcitedefaultmidpunct}
{\mcitedefaultendpunct}{\mcitedefaultseppunct}\relax
\EndOfBibitem
\bibitem{Staric:2015sta}
Belle collaboration, M.~Stari\v{c} {\em et~al.},
  \ifthenelse{\boolean{articletitles}}{\emph{{Measurement of \Dz--\Dzb mixing
  and search for \CP violation in $\Dz\to K^+K^-$, $\pi^+ \pi^−$ decays with
  the full Belle data set}},
  }{}\href{https://doi.org/10.1016/j.physletb.2015.12.025}{Phys.\ Lett.\
  \textbf{B753} (2016) 412},
  \href{http://arxiv.org/abs/1509.08266}{{\normalfont\ttfamily
  arXiv:1509.08266}}\relax
\mciteBstWouldAddEndPuncttrue
\mciteSetBstMidEndSepPunct{\mcitedefaultmidpunct}
{\mcitedefaultendpunct}{\mcitedefaultseppunct}\relax
\EndOfBibitem
\bibitem{LHCb-PAPER-2011-032}
LHCb collaboration, R.~Aaij {\em et~al.},
  \ifthenelse{\boolean{articletitles}}{\emph{{Measurement of mixing and \CP
  violation parameters in two-body charm decays}},
  }{}\href{https://doi.org/10.1007/JHEP04(2012)129}{JHEP \textbf{04} (2012)
  129}, \href{http://arxiv.org/abs/1112.4698}{{\normalfont\ttfamily
  arXiv:1112.4698}}\relax
\mciteBstWouldAddEndPuncttrue
\mciteSetBstMidEndSepPunct{\mcitedefaultmidpunct}
{\mcitedefaultendpunct}{\mcitedefaultseppunct}\relax
\EndOfBibitem
\bibitem{Blaylock:1995ay}
G.~Blaylock, A.~Seiden, and Y.~Nir,
  \ifthenelse{\boolean{articletitles}}{\emph{{The role of \CP violation in
  \Dz--\Dzb mixing}},
  }{}\href{https://doi.org/10.1016/0370-2693(95)00787-L}{Phys.\ Lett.\
  \textbf{B355} (1995) 555},
  \href{http://arxiv.org/abs/hep-ph/9504306}{{\normalfont\ttfamily
  arXiv:hep-ph/9504306}}\relax
\mciteBstWouldAddEndPuncttrue
\mciteSetBstMidEndSepPunct{\mcitedefaultmidpunct}
{\mcitedefaultendpunct}{\mcitedefaultseppunct}\relax
\EndOfBibitem
\bibitem{Bianco:2003vb}
S.~Bianco, F.~L. Fabbri, D.~Benson, and I.~Bigi,
  \ifthenelse{\boolean{articletitles}}{\emph{{A Cicerone for the physics of
  charm}}, }{}\href{https://doi.org/10.1393/ncr/i2003-10003-1}{Riv.\ Nuovo
  Cim.\  \textbf{26 n. 7-8} (2003) 1},
  \href{http://arxiv.org/abs/hep-ex/0309021}{{\normalfont\ttfamily
  arXiv:hep-ex/0309021}}\relax
\mciteBstWouldAddEndPuncttrue
\mciteSetBstMidEndSepPunct{\mcitedefaultmidpunct}
{\mcitedefaultendpunct}{\mcitedefaultseppunct}\relax
\EndOfBibitem
\bibitem{Grossman:2006jg}
Y.~Grossman, A.~L. Kagan, and Y.~Nir,
  \ifthenelse{\boolean{articletitles}}{\emph{{New physics and \CP violation in
  singly Cabibbo suppressed \D decays}},
  }{}\href{https://doi.org/10.1103/PhysRevD.75.036008}{Phys.\ Rev.\
  \textbf{D75} (2007) 036008},
  \href{http://arxiv.org/abs/hep-ph/0609178}{{\normalfont\ttfamily
  arXiv:hep-ph/0609178}}\relax
\mciteBstWouldAddEndPuncttrue
\mciteSetBstMidEndSepPunct{\mcitedefaultmidpunct}
{\mcitedefaultendpunct}{\mcitedefaultseppunct}\relax
\EndOfBibitem
\bibitem{Petrov:2006nc}
A.~A. Petrov, \ifthenelse{\boolean{articletitles}}{\emph{{Charm mixing in the
  standard model and beyond}},
  }{}\href{https://doi.org/10.1142/S0217751X06034902}{Int.\ J.\ Mod.\ Phys.\
  \textbf{A21} (2006) 5686},
  \href{http://arxiv.org/abs/hep-ph/0611361}{{\normalfont\ttfamily
  arXiv:hep-ph/0611361}}\relax
\mciteBstWouldAddEndPuncttrue
\mciteSetBstMidEndSepPunct{\mcitedefaultmidpunct}
{\mcitedefaultendpunct}{\mcitedefaultseppunct}\relax
\EndOfBibitem
\bibitem{Golowich:2007ka}
E.~Golowich, J.~Hewett, S.~Pakvasa, and A.~A. Petrov,
  \ifthenelse{\boolean{articletitles}}{\emph{{Implications of \Dz--\Dzb mixing
  for new physics}},
  }{}\href{https://doi.org/10.1103/PhysRevD.76.095009}{Phys.\ Rev.\
  \textbf{D76} (2007) 095009},
  \href{http://arxiv.org/abs/0705.3650}{{\normalfont\ttfamily
  arXiv:0705.3650}}\relax
\mciteBstWouldAddEndPuncttrue
\mciteSetBstMidEndSepPunct{\mcitedefaultmidpunct}
{\mcitedefaultendpunct}{\mcitedefaultseppunct}\relax
\EndOfBibitem
\bibitem{Ciuchini:2007cw}
M.~Ciuchini {\em et~al.}, \ifthenelse{\boolean{articletitles}}{\emph{{\Dz--\Dzb
  mixing and new physics: General considerations and constraints on the MSSM}},
  }{}\href{https://doi.org/10.1016/j.physletb.2007.08.055}{Phys.\ Lett.\
  \textbf{B655} (2007) 162},
  \href{http://arxiv.org/abs/hep-ph/0703204}{{\normalfont\ttfamily
  arXiv:hep-ph/0703204}}\relax
\mciteBstWouldAddEndPuncttrue
\mciteSetBstMidEndSepPunct{\mcitedefaultmidpunct}
{\mcitedefaultendpunct}{\mcitedefaultseppunct}\relax
\EndOfBibitem
\bibitem{Sakharov:1967dj}
A.~D. Sakharov, \ifthenelse{\boolean{articletitles}}{\emph{{Violation of \CP
  invariance, $C$ asymmetry, and baryon asymmetry of the universe}},
  }{}\href{https://doi.org/10.1070/PU1991v034n05ABEH002497}{Pisma Zh.\ Eksp.\
  Teor.\ Fiz.\  \textbf{5} (1967) 32}, Usp. Fiz. Nauk {\bf 161} (1991) 61\relax
\mciteBstWouldAddEndPuncttrue
\mciteSetBstMidEndSepPunct{\mcitedefaultmidpunct}
{\mcitedefaultendpunct}{\mcitedefaultseppunct}\relax
\EndOfBibitem
\bibitem{Huet:1994jb}
P.~Huet and E.~Sather, \ifthenelse{\boolean{articletitles}}{\emph{{Electroweak
  baryogenesis and standard model \CP violation}},
  }{}\href{https://doi.org/10.1103/PhysRevD.51.379}{Phys.\ Rev.\  \textbf{D51}
  (1995) 379}, \href{http://arxiv.org/abs/hep-ph/9404302}{{\normalfont\ttfamily
  arXiv:hep-ph/9404302}}\relax
\mciteBstWouldAddEndPuncttrue
\mciteSetBstMidEndSepPunct{\mcitedefaultmidpunct}
{\mcitedefaultendpunct}{\mcitedefaultseppunct}\relax
\EndOfBibitem
\bibitem{Alves:2008zz}
LHCb collaboration, A.~A. Alves~Jr.\ {\em et~al.},
  \ifthenelse{\boolean{articletitles}}{\emph{{The \lhcb detector at the LHC}},
  }{}\href{https://doi.org/10.1088/1748-0221/3/08/S08005}{JINST \textbf{3}
  (2008) S08005}\relax
\mciteBstWouldAddEndPuncttrue
\mciteSetBstMidEndSepPunct{\mcitedefaultmidpunct}
{\mcitedefaultendpunct}{\mcitedefaultseppunct}\relax
\EndOfBibitem
\bibitem{LHCb-DP-2014-002}
LHCb collaboration, R.~Aaij {\em et~al.},
  \ifthenelse{\boolean{articletitles}}{\emph{{LHCb detector performance}},
  }{}\href{https://doi.org/10.1142/S0217751X15300227}{Int.\ J.\ Mod.\ Phys.\
  \textbf{A30} (2015) 1530022},
  \href{http://arxiv.org/abs/1412.6352}{{\normalfont\ttfamily
  arXiv:1412.6352}}\relax
\mciteBstWouldAddEndPuncttrue
\mciteSetBstMidEndSepPunct{\mcitedefaultmidpunct}
{\mcitedefaultendpunct}{\mcitedefaultseppunct}\relax
\EndOfBibitem
\bibitem{LHCb-PROC-2010-056}
I.~Belyaev {\em et~al.}, \ifthenelse{\boolean{articletitles}}{\emph{{Handling
  of the generation of primary events in Gauss, the LHCb simulation
  framework}}, }{}\href{https://doi.org/10.1088/1742-6596/331/3/032047}{J.\
  Phys.\ Conf.\ Ser.\  \textbf{331} (2011) 032047}\relax
\mciteBstWouldAddEndPuncttrue
\mciteSetBstMidEndSepPunct{\mcitedefaultmidpunct}
{\mcitedefaultendpunct}{\mcitedefaultseppunct}\relax
\EndOfBibitem
\bibitem{LHCb-PROC-2011-006}
M.~Clemencic {\em et~al.}, \ifthenelse{\boolean{articletitles}}{\emph{{The
  \lhcb simulation application, Gauss: Design, evolution and experience}},
  }{}\href{https://doi.org/10.1088/1742-6596/331/3/032023}{J.\ Phys.\ Conf.\
  Ser.\  \textbf{331} (2011) 032023}\relax
\mciteBstWouldAddEndPuncttrue
\mciteSetBstMidEndSepPunct{\mcitedefaultmidpunct}
{\mcitedefaultendpunct}{\mcitedefaultseppunct}\relax
\EndOfBibitem
\bibitem{LHCb-DP-2018-004}
D.~M{\"u}ller, M.~Clemencic, G.~Corti, and M.~Gersabeck,
  \ifthenelse{\boolean{articletitles}}{\emph{{ReDecay: A novel approach to
  speed up the simulation at LHCb}},
  }{}\href{https://doi.org/10.1140/epjc/s10052-018-6469-6}{Eur.\ Phys.\ J.\
  \textbf{C78} ({2018}) 1009},
  \href{http://arxiv.org/abs/1810.10362}{{\normalfont\ttfamily
  arXiv:1810.10362}}\relax
\mciteBstWouldAddEndPuncttrue
\mciteSetBstMidEndSepPunct{\mcitedefaultmidpunct}
{\mcitedefaultendpunct}{\mcitedefaultseppunct}\relax
\EndOfBibitem
\bibitem{LHCb-DP-2012-004}
R.~Aaij {\em et~al.}, \ifthenelse{\boolean{articletitles}}{\emph{{The \lhcb
  trigger and its performance in 2011}},
  }{}\href{https://doi.org/10.1088/1748-0221/8/04/P04022}{JINST \textbf{8}
  (2013) P04022}, \href{http://arxiv.org/abs/1211.3055}{{\normalfont\ttfamily
  arXiv:1211.3055}}\relax
\mciteBstWouldAddEndPuncttrue
\mciteSetBstMidEndSepPunct{\mcitedefaultmidpunct}
{\mcitedefaultendpunct}{\mcitedefaultseppunct}\relax
\EndOfBibitem
\bibitem{LHCb-PAPER-2014-013}
LHCb collaboration, R.~Aaij {\em et~al.},
  \ifthenelse{\boolean{articletitles}}{\emph{{Measurement of \CP asymmetry in
  $\Dz\to \Km\Kp$ and $\Dz\to \pim\pip$ decays}},
  }{}\href{https://doi.org/10.1007/JHEP07(2014)041}{JHEP \textbf{07} (2014)
  041}, \href{http://arxiv.org/abs/1405.2797}{{\normalfont\ttfamily
  arXiv:1405.2797}}\relax
\mciteBstWouldAddEndPuncttrue
\mciteSetBstMidEndSepPunct{\mcitedefaultmidpunct}
{\mcitedefaultendpunct}{\mcitedefaultseppunct}\relax
\EndOfBibitem
\bibitem{LHCb-PAPER-2017-004}
LHCb collaboration, R.~Aaij {\em et~al.},
  \ifthenelse{\boolean{articletitles}}{\emph{{Measurement of $\Bs$ and $\Dsm$
  meson lifetimes}},
  }{}\href{https://doi.org/10.1103/PhysRevLett.119.101801}{Phys.\ Rev.\ Lett.\
  \textbf{119} (2017) 101801},
  \href{http://arxiv.org/abs/1705.03475}{{\normalfont\ttfamily
  arXiv:1705.03475}}\relax
\mciteBstWouldAddEndPuncttrue
\mciteSetBstMidEndSepPunct{\mcitedefaultmidpunct}
{\mcitedefaultendpunct}{\mcitedefaultseppunct}\relax
\EndOfBibitem
\bibitem{johnson}
N.~L. Johnson, \ifthenelse{\boolean{articletitles}}{\emph{{Systems of frequency
  curves generated by methods of translation}},
  }{}\href{https://doi.org/10.1093/biomet/36.1-2.149}{Biometrika \textbf{36}
  (1949) 149}\relax
\mciteBstWouldAddEndPuncttrue
\mciteSetBstMidEndSepPunct{\mcitedefaultmidpunct}
{\mcitedefaultendpunct}{\mcitedefaultseppunct}\relax
\EndOfBibitem
\end{mcitethebibliography}
